%% file: main.tex
\documentclass[10pt,conference]{IEEEtran}
\usepackage[letterpaper, left=0.63in, right=0.63in, bottom=1.05in, top=0.7in]{geometry}
\IEEEoverridecommandlockouts

\usepackage{cite}
\usepackage{amsmath}
\usepackage{graphicx}
\usepackage{amsfonts}
\usepackage{lscape}
\usepackage{multirow}
\usepackage[table,xcdraw]{xcolor}
\usepackage{booktabs}
\usepackage{tabu}
\usepackage{colortbl}
\usepackage{hhline}
\usepackage[ruled,vlined]{algorithm2e}
\usepackage{algpseudocode}
\usepackage{tikz}
\usetikzlibrary{decorations.pathreplacing,calc}
\usepackage{adjustbox}

\DeclareMathOperator*{\argmin}{arg\,min}
\newcommand{\bs}[1]{\boldsymbol{#1}}

\newcommand{\mb}[1]{\mathbf{#1}}

\newcommand{\tikzmark}[1]{\tikz[overlay,remember picture] \node (#1) {};}
\input{acronym_list}
\begin{document}

\title{A Foundation Model for Massive MIMO Precoding with an Adaptive Per-User Rate-Power Tradeoff

}
\author{\IEEEauthorblockN{Jérôme~Emery, Ali~Hasanzadeh~Karkan, Jean-François~Frigon, and François~Leduc-Primeau\\
\\
\footnotesize{Department of Electrical Engineering, Polytechnique Montreal, Montreal, QC H3T 1J4, Canada.}\\
\footnotesize{Emails: \{jerome.emery, ali.hasanzadeh-karkan, j-f.frigon, francois.leduc-primeau\}@polymtl.ca.} \\
}}

\maketitle

\input{Abstract}

\input{Introduction}

\input{Sysmodel}

\input{FoundationModel}

\input{Results}

\input{Conclusion}

\section*{Acknowledgement}
This work was supported by Ericsson - Global Artificial Intelligence Accelerator AI-Hub Canada in Montr\'{e}al and jointly funded by NSERC Alliance Grant 566589-21 (Ericsson, ECCC, Innov\'{E}\'{E}).

\bibliographystyle{IEEEtran}
\bibliography{main}

\end{document}

%% file: acronym_list.tex
\usepackage[acronym]{glossaries}
\newacronym{BS}{BS}{base station}
\newacronym{PS}{PS}{phase-shifter}
\newacronym{RL}{RL}{reinforcement learning}
\newacronym{AP}{AP}{analog precoder}
\newacronym{FC-HBF}{FC-HBF}{fully-connected HBF}
\newacronym{FSA-HBF}{FSA-HBF}{fixed subarray HBF}
\newacronym{DSA-HBF}{DSA-HBF}{dynamic subarray HBF}
\newacronym{BF}{BF}{beamforming}
\newacronym{UE}{UE}{user equipment}
\newacronym{AWGN}{AWGN}{additive white gaussian noise}
\newacronym{MIMO}{MIMO}{multiple-input multiple-output}
\newacronym{MISO}{MISO}{multiple-input single-output}
\newacronym{RF}{RF}{radio frequency}
\newacronym{RIS}{RIS}{reconfigurable intelligent surfaces}
\newacronym{IOT}{IOT}{internet-of-things}
\newacronym{CL}{CL}{convolutional layer}
\newacronym{FDD}{FDD}{frequency division duplex}
\newacronym{TDD}{TDD}{time division duplex}
\newacronym{CSI}{CSI}{channel state information}
\newacronym{DNN}{DNN}{deep neural network}
\newacronym{DP}{DP}{digital precoder}
\newacronym{DL}{DL}{deep learning}
\newacronym{SVD}{SVD}{singular-value decomposition}
\newacronym{CNN}{CNN}{convolution neural network}
\newacronym{FDP}{FDP}{fully digital precoder}
\newacronym{SE}{SE}{spectral efficiency}
\newacronym{OFDM}{OFDM}{orthogonal frequency division multiplexing}
\newacronym{OMP}{OMP}{orthogonal matching pursuit}
\newacronym{FL}{FL}{fully-connected layer}
\newacronym{HSHO}{HSHO}{Hybrid Structured Heuristic Optimization}
\newacronym{HBF}{HBF}{hybrid beamforming}
\newacronym{IA}{IA}{initial access}
\newacronym{mm-Wave}{mm-Wave}{millimeter wave}
\newacronym{mMIMO}{mMIMO}{massive MIMO}
\newacronym{SINR}{SINR}{signal-to-interference-plus-noise ratio}
\newacronym{SNR}{SNR}{signal-to-noise ratio}
\newacronym{RSSI}{RSSI}{received signal strength indicator}
\newacronym{PZF}{PZF}{phase zero forcing}
\newacronym{PSO}{PSO}{particle swarm optimization}
\newacronym{ZF}{ZF}{zero forcing}
\newacronym{O-FDP}{O-FDP}{optimal fully digital precoder}
\newacronym{JT}{JT}{joint transmission}
\newacronym{CU}{CU}{central unit}
\newacronym{MSE}{MSE}{mean square error}
\newacronym{CEL}{CEL}{cross entropy loss}
\newacronym{CB}{CB}{conjugate beamforming}
\newacronym{NC}{NC}{network controller}
\newacronym{CoMP}{CoMP}{coordinated multi point}
\newacronym{CF-mMIMO}{CF-mMIMO}{cell-free massive MIMO}
\newacronym{CF-HBF}{CF-HBF}{cell-free hybrid beamforming}
\newacronym{CF-BF}{CF-BF}{cell-free beamforming}
\newacronym{MLDG}{MLDG}{meta-learning domain generalization}
\newacronym{MAML}{MAML}{model-agnostic meta-learning}
\newacronym{WSR}{WSR}{weighted sum rate}
\newacronym{SSL}{SSL}{self-supervised learning}
\newacronym{WMMSE}{WMMSE}{weighted minimum mean squared error}
\newacronym{LOS}{LOS}{line-of-sight}
\newacronym{NLOS}{NLOS}{non line-of-sight}
\newacronym{FLOPs}{FLOPs}{floating point operations}
\newacronym{EE}{EE}{energy efficiency}
\newacronym{FFN}{FFN}{feedforward network}

%% file: Abstract.tex
\begin{abstract}

Deep learning (DL) has emerged as a solution for precoding in massive multiple-input multiple-output (mMIMO) systems due to its capacity to learn the characteristics of the propagation environment. However, training such a model requires high-quality, local datasets at the deployment site, which are often difficult to collect. We propose a transformer-based foundation model for mMIMO precoding that seeks to minimize the energy consumption of the transmitter while dynamically adapting to per-user rate requirements. At equal energy consumption, zero-shot deployment of the proposed foundation model significantly outperforms zero forcing, and approaches weighted minimum mean squared error performance with $8\times$ less complexity. To address model adaptation in data-scarce settings, we introduce a data augmentation method that finds training samples similar to the target distribution by computing the cosine similarity between the outputs of the pre-trained feature extractor. Our work enables the implementation of DL-based solutions in practice by addressing challenges of data availability and training complexity. Moreover, the ability to dynamically configure per-user rate requirements can be leveraged by higher level resource allocation and scheduling algorithms for greater control over energy efficiency, spectral efficiency and fairness.

\end{abstract}

%% file: Introduction.tex
\section{Introduction}\label{sec:intro}

Introduced at scale in the fifth generation of wireless communication, \gls{mMIMO} technology has been instrumental in achieving higher \gls{SE} and capacity in wireless communication networks. By leveraging multiple antennas at the transmit and/or receive side, multiple data streams can be transmitted on the same time-frequency resource while limiting inter-user interference \cite{iui}. However, the performance of this technology relies on the design of a precoder before every transmission, which requires solving a complex, non-convex optimization problem in real-time.

\Gls{DL} methods have been proposed to compute precoders for \gls{mMIMO} systems to address the complexity of real-time optimization \cite{us_dl_mm_bf, slmp, low_complexity_dl, unsupervised_graph}. By learning the mapping between \gls{CSI} and a precoder, \gls{DL}-based methods significantly reduce computational complexity compared to traditional optimization. However, deploying a \gls{DNN} requires extensive training and testing on a local high-quality dataset that is rarely available in practice.

Recent work has explored meta-learning and transfer learning to adapt models to new environments using few data samples. The authors of \cite{maml_hbf} implement \gls{MAML} to enable rapid adaptation of \gls{DL} precoding models, but only to generalize with respect to the distribution of users. In \cite{mltl}, two offline training algorithms as well as an online adaptation method based on transfer learning and meta-learning are proposed. A similar algorithm is proposed in \cite{sage} to train a model on statistical data, and to deploy it to a realistic dataset produced by ray-tracing simulation. Although these techniques are promising, the amount of data needed to adapt the model is still large and their performance when deployed without adaptation is lower than that of a simple \gls{ZF} precoder.

Emerging from a new \gls{DL} paradigm, foundation models trained on extensive datasets provide excellent generalization across a variety of domains. Capitalizing on the reduction of training complexity and data dependence, foundation models for the physical layer have been proposed. A model based on the transformer architecture is presented in \cite{catak2025bert4mimofoundationmodelusing} for \gls{CSI} reconstruction. The pre-trained model is evaluated on various mobility scenarios and channel conditions, confirming its generalization capabilities. Similarly, for a \gls{CSI} feedback task, \cite{foundation_2} considers pre-training, then adapting to reduce both the complexity of training and the need for local datasets.  

Another line of work has explored energy-efficient \gls{mMIMO} precoding techniques such as hybrid beamforming and antenna selection. \gls{DL} presents itself as a powerful method to efficiently find approximate solutions to these complex decision problems. Similar to our work, the authors of \cite{noisy_csi} maximize the \gls{EE} of a \gls{mMIMO} system, by minimizing the number of active antennas used for transmission, subject to an average sum-rate constraint. The \gls{DNN} achieves significant energy gains, but it is trained to satisfy a fixed sum-rate, limiting its generalization capability to time-varying \gls{SE} requirements. 

In this work, we address limitations in generalization and robustness of \gls{DL}-based precoding by proposing a foundation model for \gls{mMIMO} precoding. We first describe how to train a \gls{DNN} on a large dataset with a focus on training a robust, general feature extractor. We then discuss adaptation to new sites with no or few data samples, which we refer to as the \emph{zero-shot} and \emph{few-shot} settings respectively. We also provide a data augmentation method that finds training environments similar to the deployment site by computing a similarity metric over the outputs of the feature extractor. Going beyond \gls{SE} maximization, we formulate a flexible training objective that seeks to minimize energy consumption, while satisfying dynamic per-user rate requirements. As such, our model adapts to varying service demands by reducing unnecessary energy consumption of the transmitter.

%% file: Sysmodel.tex
\section{System Model}\label{sec:sysmodel}
\subsection{Communication Model}
We study a multi-user \gls{mMIMO} system where a \gls{BS} is configured with $N_T$ transmit antennas. The \gls{BS} concurrently serves $N_U$ users, each equipped with a single antenna. We consider a \gls{FDP} at the \gls{BS} for the downlink transmission expressed as $\mathbf{W} = \left [ \mathbf{w}_1, \hdots, \mathbf{w}_u, \hdots, \mathbf{w}_{N_U} \right ] \in \mathbb{C}^{N_T \times N_U}$, normalized to satisfy the maximum power constraint. To make explicit the effect of the transmit power on the rate, we multiply $\mb{W}$ by a scaling factor $\gamma \in \left [ 0, 1\right ]$. Let $\mathbf{x} = \left [ x_1, \hdots, x_u, \hdots, x_{N_U}\right ]$ contain the independent transmitted symbols for all users, normalized such that $\mathbb{E}\left [ \mathbf{x}\mathbf{x}^{\dagger} \right ] = \frac{1}{N_u} \mathbb{I}_{N_u}$, and let $\mathbf{h}_u \in \mathbb{C}^{N_T \times 1}$ be the channel vector from the $N_T$ \gls{BS} antennas to user $u$. The signal received by each user can be expressed as
\begin{equation}
    \mathbf{y}_u =  \mathbf{h}_{u}^{\dagger} \sum_{\forall u}  \sqrt{\gamma} \cdot \mathbf{w}_{u} x_u + \bs{\eta}\, ,
\end{equation}
 where $\bs{\eta}$ is the additive white Gaussian noise with power $\sigma^2$. The \gls{SINR} for user $u$ is given by
\begin{equation}
    \text{SINR}(\sqrt{\gamma} \cdot\mb{w}_{u}) = \frac{ \big|\sqrt{\gamma} \cdot\mb{h}^{\dagger}_{u} \mb{w}_{u} \big|^2}{\sum_{j \neq u} \big|\sqrt{\gamma} \cdot\mb{h}^{\dagger}_{u} \mb{w}_{j} \big|^2 + \sigma^2}\,,
\end{equation}
and for a precoding vector $\mathbf{w_u}$, the rate for user $u$ is
\begin{equation}\label{eq:sum-rate}
    R(\sqrt\gamma \cdot \mb{w_u}) = \text{log}_2 \Bigl(  1+ \text{SINR}(\sqrt \gamma \cdot \mb{w}_{u}) \Bigr).
\end{equation}

\subsection{Energy Model}\label{sec:energy}
In a \gls{FDP}, each antenna is connected to a dedicated \gls{RF} chain, providing great beamforming flexibility by enabling per-antenna amplitude and phase control. However, this comes at the cost of increased energy consumption due to the large number of active \gls{RF} circuits. Dynamic antenna selection is a promising technique for reducing the energy cost of \gls{mMIMO} systems. 

In our energy model, we assume each antenna can be omitted for one transmission. Let $P_{RF}$ be the fixed energy cost for turning on one antenna. Given a binary mask $\mb{\Omega} = [ \omega_1, \omega_2, \hdots, \omega_{N_T}]$ where $\omega_i = 1$ indicates turning on antenna $i$, we can express the fixed energy consumption of the active antennas as $P_{RF} \cdot \sum_{i=1}^{N_T}\omega_i$. For greater control on the energy, we reduce the maximum transmit power $P_{TX}$ by learning the scaling factor $\gamma$.
Summing the transmit power and the antenna selection costs, the total energy consumption can be written as 
\begin{equation}\label{eq:EC}
    E\bigl ( \mb{\Omega}, \gamma\bigr ) = \gamma \cdot P_{TX} + P_{RF}\sum_{i=1}^{N_T}\omega_i \,.
\end{equation}
In this formulation, we control the global transmit power $P_{TX}$ to facilitate the comparison with conventional baselines. However, the proposed approach can be easily modified to consider per-antenna power constraints. 

\subsection{Problem Formulation}\label{sec:problem}

Given per-user \gls{SE} requirements $R_{u}^\star$ for each user $u$, the objective is to minimize the energy consumption $E$. This problem can be formulated as 
\begin{equation}\label{eq:opt}
    \begin{aligned}
    \underset{\mb{W}, \mb{\Omega}, \gamma}{\min} \quad & E(\mb{\Omega}, \gamma) \\
    \text{s.t.} \quad & \sum_{\forall u} \mb{w}_{u}^{\dagger} \mb{w}_{u} \leq P_{TX} \,, \\
    \quad & R(\mb{\gamma \cdot \Omega} \odot \mb{w}_u) \geq R_{u}^\star, \quad \forall u \,, \\
    \quad & \omega_i \in \{0, 1\}, \quad \forall i \,,\\
    \quad & \gamma \in \bigl [0, 1 \bigr ]\,,
\end{aligned}
\end{equation}
where $\odot$ denotes the element-wise product of two vectors. The formulated optimization problem is difficult to solve analytically due to mixed binary and continuous variables as well as the dependence between the objective and the constraints through $\mb{\Omega}$ and $\gamma$. We train a \gls{DNN} to learn this complex task.

\subsection{Baselines}
We consider two classical baseline algorithms for fully digital precoding, each providing a trade-off between computational complexity and performance.
\begin{itemize}
    \item \gls{ZF} consists of simply pre-multiplying the transmit data with the pseudo-inverse of the channel matrix. It has been shown that \gls{ZF} is close to optimal in high \gls{SNR} scenarios, but its performance quickly degrades as noise increases \cite{zf}.
    \item \Gls{WMMSE} provides near-optimal performance by converting throughput maximization into a mean-squared error minimization problem, and solving it iteratively \cite{shi_wmmse}. However, computing the \gls{WMMSE} precoder comes at the cost of high computational complexity due to the iterative optimization steps.
\end{itemize}

\subsection{Dataset Generation}

To generate training and evaluation data, we generate \gls{CSI} datasets from custom \gls{BS} configurations, using MATLAB's ray-tracing toolbox and OpenStreetMap~\cite{OpenStreetMap} to simulate realistic propagation conditions based on 3D models of outdoor environments at specified coordinates. 

We consider \gls{BS}s equipped with a uniform planar array antenna consisting of $8 \times 8$ elements. The antenna operates at a carrier frequency of 2 GHz, and the transmit power is set to 20 watts. Additionally, we account for a system loss of 10 dB. Note that our simulations support the modeling of up to ten reflections. 
Each channel sample is generated by placing a \gls{UE} around the \gls{BS} and simulating its \gls{CSI}. From a collection of single-user channel realizations, we construct multi-user \gls{CSI}s by randomly sampling $N_U$ users and forming a \gls{CSI} matrix $\mb{H} \in \mathbb{C}^{N_U \times N_T}$. 

Our \gls{CSI} training dataset consists of a collection of datasets corresponding to different environments, to encourage the model to learn general representations. We generate $n$ samples of \gls{CSI} data $\{ \mathbf{H}_e^{(j)} \}_{j=1}^{n}$ for various environments $e$. Furthermore, to group together samples achieving similar sum-rates, we split each environment into \gls{LOS} and \gls{NLOS} users. In total, $k$ datasets are generated, with $k$ being twice the amount of training environments considered. 

\subsection{Addressing Robustness and Generalization} \label{sec:generalization}

\Gls{SSL} for \gls{mMIMO} precoding performs well when training and deployment occur in the same environment. However, the need for a high-quality local dataset limits practical deployment. Intuitively, we would expect that a model trained at one site would generalize. Yet, Table~\ref{tab:cross_performance} shows a substantial performance drop when deploying on unseen environments.

We hypothesize that this is due to the model overfitting to spurious features that are predictive on the training distribution, but not causally linked to the task. Such feature reliance often yields strong in-distribution performance but poor generalization to out-of-distribution data.

These results highlight the need for more robust DL-based precoding. If the model has learned spurious features over a local dataset, then sudden environmental changes at the base station could immediately result in a significant performance drop. This experiment motivates our work to train robust, generalizable precoding models with strong zero-shot and few-shot performance.

\renewcommand{\arraystretch}{1}
\begin{table}[t]
\caption{Impact on sum-rate (b/s/Hz) from training a model on one site and deploying to another $(-\textnormal{ log }\sigma^2 = 13)$.}
\label{tab:cross_performance}
\resizebox{\columnwidth}{!}{%
\begin{tabular}{lc|cccccc|}
\cline{3-8}
 &
  \multicolumn{1}{l|}{} &
  \multicolumn{6}{c|}{\textbf{Deployment}} \\ \cline{3-8} 
 &
  \textbf{Environment} &
  \begin{tabular}[c]{@{}c@{}}UdeM\\ LOS\end{tabular} &
  \begin{tabular}[c]{@{}c@{}}UdeM\\ NLOS\end{tabular} &
  \begin{tabular}[c]{@{}c@{}}Old port\\ LOS\end{tabular} &
  \begin{tabular}[c]{@{}c@{}}Old port\\ NLOS\end{tabular} &
  \begin{tabular}[c]{@{}c@{}}Oka park\\ LOS\end{tabular} &
  \begin{tabular}[c]{@{}c@{}}Oka park\\ NLOS\end{tabular} \\ \hline
\multicolumn{1}{|l|}{} &
  UdeM LOS &
  \cellcolor[HTML]{AAE1AB}32.36 &
  \cellcolor[HTML]{EFE8A7}14.82 &
  \cellcolor[HTML]{FFCDD2}12.63 &
  \cellcolor[HTML]{FFCDD2}8.34 &
  \cellcolor[HTML]{EFE8A7}29.27 &
  \cellcolor[HTML]{FFCDD2}9.48 \\
\multicolumn{1}{|l|}{} &
  UdeM NLOS &
  \cellcolor[HTML]{EFE8A7}26.16 &
  \cellcolor[HTML]{AAE1AB}17.24 &
  \cellcolor[HTML]{FFCDD2}11.76 &
  \cellcolor[HTML]{FFCDD2}8.88 &
  \cellcolor[HTML]{EFE8A7}27.23 &
  \cellcolor[HTML]{FFCDD2}11.71 \\
\multicolumn{1}{|l|}{} &
  Old port LOS &
  \cellcolor[HTML]{FFCDD2}13.02 &
  \cellcolor[HTML]{FFCDD2}7.81 &
  \cellcolor[HTML]{AAE1AB}34.13 &
  \cellcolor[HTML]{FFCDD2}12.90 &
  \cellcolor[HTML]{FFCDD2}13.08 &
  \cellcolor[HTML]{FFCDD2}3.37 \\
\multicolumn{1}{|l|}{} &
  Old port NLOS &
  \cellcolor[HTML]{FFCDD2}15.56 &
  \cellcolor[HTML]{FFCDD2}11.18 &
  \cellcolor[HTML]{EFE8A7}27.80 &
  \cellcolor[HTML]{AAE1AB}19.23 &
  \cellcolor[HTML]{FFCDD2}18.29 &
  \cellcolor[HTML]{FFCDD2}8.54 \\
\multicolumn{1}{|l|}{} &
  Oka park LOS &
  \cellcolor[HTML]{EFE8A7}26.88 &
  \cellcolor[HTML]{FFCDD2}13.97 &
  \cellcolor[HTML]{FFCDD2}12.78 &
  \cellcolor[HTML]{FFCDD2}8.56 &
  \cellcolor[HTML]{AAE1AB}33.78 &
  \cellcolor[HTML]{FFCDD2}10.45 \\
\multicolumn{1}{|l|}{\multirow{-6}{*}{\rotatebox{90}{\textbf{Training}}}} &
  Oka park NLOS &
  \cellcolor[HTML]{FFCDD2}8.75 &
  \cellcolor[HTML]{FFCDD2}6.11 &
  \cellcolor[HTML]{FFCDD2}6.32 &
  \cellcolor[HTML]{FFCDD2}4.50 &
  \cellcolor[HTML]{FFCDD2}9.78 &
  \cellcolor[HTML]{AAE1AB}16.54 \\ \hline
\end{tabular}%
}
\end{table} 

%% file: FoundationModel.tex
\section{Foundation model} \label{sec:foundation}
In this section, we present an architecture and an algorithm to train a foundation model for \gls{mMIMO} precoding with an adaptive per-user rate-power tradeoff. The focus is on learning robust representations transferable to new deployment sites, while also being able to reduce the \gls{BS} power consumption when rate requirements are below the maximum achievable.

\subsection{Training Objective}\label{sec:train_obj}
We begin by formulating \eqref{eq:opt} into a multi-objective differentiable loss function. To satisfy the user-rate demands, we convert the hard constraints into a soft approximation by minimizing the mean-squared error between actual user rates and the corresponding user-rate requirement $R_{u}^\star$. The adaptive rate objective is given by:
\begin{equation}\label{eq:AR}
    \mathcal{L}_{AR} = \frac{1}{N_U}\sum_{u} \Bigr( R(\gamma \cdot \mb{\Omega} \odot\mb{w}_u) - R_{u}^\star \Bigl)^2.
\end{equation}

The energy consumption objective, $\mathcal{L}_{\text{EC}}$, is simply given by $E(\mb{\Omega}, \gamma)$. Summing these two objectives yields our loss function 
\begin{equation}\label{eq:moo}
\mathcal{L} = \mu \mathcal{L}_{\text{AR}} + (1-\mu)\mathcal{L}_{\text{EC}}\,,
\end{equation}
where $\mu$ is a hyper-parameter tuned to balance energy consumption and user-rate satisfaction. In practice, we set $\mu$ large enough such that energy consumption is only reduced when user-rates are satisfied. Note that the formulated training loss is only a function of the \gls{CSI}, rate requirements, and the predicted outputs. Thus we can train the model through \gls{SSL}, by minimizing $\mathcal{L}$ without the need for pre-computed labels. 

\begin{figure}[t]
    \centering
    \includegraphics[width=0.8\columnwidth]{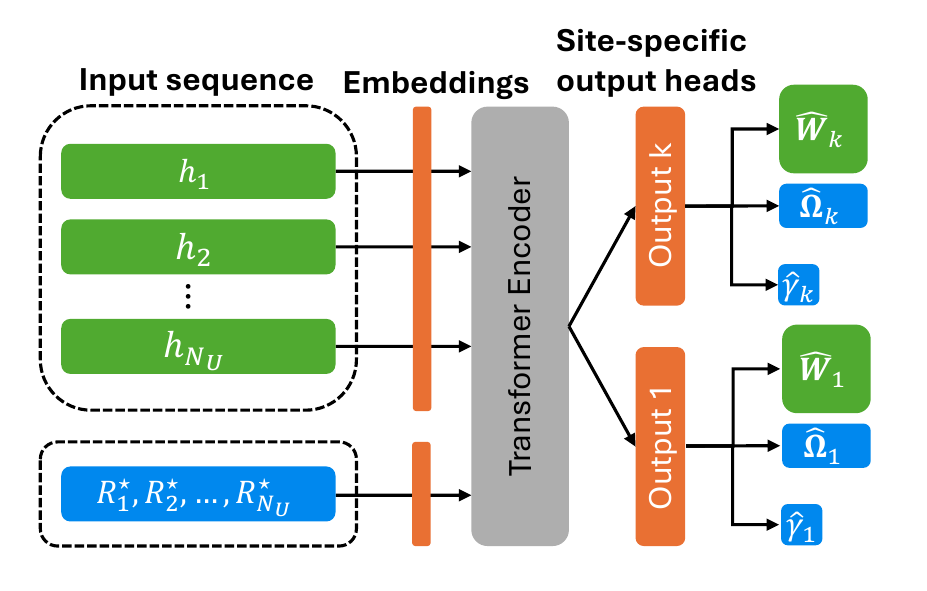}
    \caption{\gls{DNN} architecture of the proposed foundation model.}
    \label{fig:arch}
\end{figure}

\subsection{Foundation Model}

\subsubsection{Transformer-encoder architecture} Our model adapts a transformer-encoder architecture, which leverages the self-attention mechanism to compute dependencies between input tokens. To align our input, the \gls{CSI} matrix $\mathbf{H} \in \mathbb{C}^{N_T \times N_U}$, to the sequential-based inputs of a transformer encoder, we deconstruct $\mathbf{H}$ into $N_U$ single-user \gls{CSI} vectors $\mb{h}_u \in \mathbb{C}^{N_T \times 1}$. Each complex vector is then flattened into a real-valued vector $\bigl[\Re(\mb{h}_u), \Im(\mb{h}_u)\bigr]$ by concatenating the real and imaginary parts. This feature representation allows us to input sequences of tokens into the model. Each \gls{CSI} token is then projected through a linear embedding layer. 

The user-rate requirements to be satisfied are appended to the input sequence as an extra context token $\mb{R}^\star = \bigl[R_1^\star, R_2^\star, \hdots, R_{N_U}^{\star} \bigr]$. This token is assigned its own embedding layer to encode a representation distinct from the \gls{CSI}.

One transformer block is comprised of a self-attention mechanism followed by a \gls{FFN}. Self-attention learns dependencies and relationships between the \gls{CSI} tokens by computing attention scores pairwise between each input. Intuitively, the spatial relationship between users is needed to learn the influence of user positions on inter-user interference. The attention outputs are then passed through the \gls{FFN} along with the original tokens. This structure forms one transformer block, which can be repeated sequentially to increase capacity. To form the precoding matrix, antenna selection vector, and power scaling factor, we append two parallel output layers after the final transformer block. The first layer maps the \gls{CSI} features to the precoding matrix. The second layer projects the features to a vector of length $N_T + 1$ before applying a sigmoid function to constrain the values between $\bigl[ 0, 1 \bigr ]$. The last element of this vector is the power scaling factor, while the remaining elements form the antenna selection. Given the discrete nature of the threshold function, we use the straight-through estimator which approximates the gradient of the threshold function with the gradient of the preceding sigmoid.

\subsubsection{Multi-head output} To train the foundation model, we assign a distinct output head to every training environment. During the forward pass, data samples flow through the feature extractor, before branching to their respective output layers. Formally, let $\left [ \mb{\widehat{W}}, \mb{\widehat{\Omega}}, \widehat{\gamma}\right ] = \Phi_{\mathbf{\theta}}\left ( \mathbf{H}, \mb{R^{\star}} \right )$ be a mapping from the \gls{CSI} matrix and the rate requirements to the precoding outputs. We separate our network into a feature extractor $f_{\mathbf{\theta}}( \cdot )$ and the site-specific output layers $ \mathbf{Z} = \left [ \mb{Z}_1, \hdots, \mb{Z}_k \right ]$ where $\mb{Z}_e$ represents both the precoding output layer $Z_e^{\mb{W}}$ and the energy consumption output layer $Z_e^{\mb{\Omega}, \gamma}$ for one environment. The training architecture can be viewed as a collection of $k$ distinct DNNs expressed as 
$$
\Phi_{\mathbf{\theta}}^{e}(\mathbf{H}, \mb{R}^\star) = \mb{Z}_ef_{\theta}(\mathbf{H}, \mb{R}^\star) \, ,
$$ 
such that each model shares the same weights except for the output layers. Considering multiple output layers, the training objective becomes 
$$ 
\argmin_{\theta, \mathbf{Z}} \sum_{e=1}^k \sum_{j=1}^{n} \mathcal{L}(\mb{Z}_e f_{\theta}(\mathbf{H}^{(j)}, \mb{R}_j^\star))
$$ 
as the model is learning the weights of the feature extractor along with the site-specific output layers.

This formulation is key to our proposed method. We allow the models to adapt to each environment, constrained to a shared feature extractor. This improves the robustness and generalization as the learned representations are common to all environments. At deployment time, we discard the site-specific output heads and only transfer the feature extractor.

\subsubsection{Pre-training}

Instead of directly minimizing $\mathcal{L}$, we first train the foundation model to maximize the sum-rate $R(\mb{W})$. This step mitigates convergence issues from directly learning multiple objectives. During the initial pre-training phase, we implement a \textit{Site-aware weighted gradient descent} algorithm aimed at normalizing the update steps based on the site-specific upper-bound sum-rates. We find through our experiments that naively maximizing the average sum-rates disproportionately favors high sum-rate environments and results in mediocre performance for inputs with lower \gls{SINR}. Our weighted gradient descent algorithm addresses this issue by accounting for the differing upper bound sum-rates among our training samples. The pre-training procedure is detailed in the first part of Algorithm~\ref{Algorithm_1}.

Before training begins, we initialize the upper bound sum-rate $R_e^{\max}$ for each training environment with the \gls{WMMSE} algorithm. For each iteration, the gradient weight is calculated as $\alpha_e = (R_e-R_e^{\max})^2$ where $R_e$ is the sum-rate for environment $e$ from the previous iteration. We then clamp the weights vector to values between $1$ and $T$, to prevent extreme gradient updates in the first iterations, and to allow continued training for environments that surpass the \gls{WMMSE} bound.  
Finally, the weights are normalized. After the forward pass, the gradients are computed for each output layer, and summed with the gradient weights.

\subsubsection{Multi-objective training}

Starting from the initialized model, we switch to the multi-objective training objective \eqref{eq:moo}. This phase of the foundation model training is described by the second part of Algorithm~\ref{Algorithm_1}. The model learns to predict $\mb{W}$ such that the user-rates match the rate requirements fed into the model as inputs. The antenna selection and power scaling are also jointly predicted. To simulate the user-rate requirements for one training iteration, we generate $N_U$ random rate requirements $R_u^\star \sim \text{Unif}(0, 1)$, scaled and normalized to sum to $\beta \cdot R_e^{\max}$ where $\beta\sim \text{Unif}(0, 1)$, and $R_e^{\max}$ is the average maximum sum-rate for environment $e$ computed with \gls{WMMSE}.

\begin{algorithm}[t]
\caption{Foundation model training}
\label{Algorithm_1}
\small
\SetKwInOut{Input}{Input}
\SetKwInOut{Initialize}{Initialize}
\Input{$B$ CSI batches for $k$ training environments: $\left [ \left\{\mb{H}_1^{(b)}\right\}_{b=1}^B, \left\{\mb{H}_2^{(b)}\right\}_{b=1}^B, \hdots, \left\{\mb{H}_k^{(b)}\right\}_{b=1}^B \right ]$}
\Initialize{Upper bound sum-rates $\left [ R_1^{\max} \hdots R_k^{\max}\right ]$, learning rate $\lambda$, parameters $\mathbf{\theta}$, threshold $T$}

\tikzmark{top1}
\For{$t \leftarrow 1$ \KwTo Epochs}{
\For{$b \leftarrow 1$ \KwTo $B$}{
    \For{$e \leftarrow 1$ \KwTo $k$}{
        Forward pass: $ \widehat{\mathbf{W}}_e \leftarrow \Phi_{\mathbf{\theta}}^{e}(\mathbf{H}_e^{(b)})$ \\
        Sum-rate: $R_e \leftarrow -R(\widehat{\mathbf{W}}_e)$\\
        Environment weight: $\alpha_e = (R_e - R_e^{\max})^2$ \tikzmark{label1} \\
        Clamp weight: $\alpha_e \leftarrow \min(T, \max(1, \alpha_e))$\\
    }
    Normalize environment weights: $\alpha_e \leftarrow \frac{\alpha_e}{\sum_{j=1}^{k} \alpha_j}$  \\
    Update parameters: 
    $\theta \leftarrow \theta - \lambda \sum_{e=1}^{k} \alpha_e \nabla (-\nabla R(\widehat{\mb{W}_e}))$\\
} \tikzmark{bottom1} 
}
\tikzmark{top}
\For{$t \leftarrow 1$ \KwTo Epochs}{
\For{$b \leftarrow 1$ \KwTo $B$}{
    User-rate requirements $R_u^\star \sim \text{Unif}(0, 1) \, , \forall u$ \\
    Random sum-rate scaling $\beta \sim \text{Unif}(0, 1)$ \\
    \For{$e \leftarrow 1$ \KwTo $k$}{
        $R_u^\star \leftarrow \beta \cdot R_e^{\max}\frac{R_u^\star}{\sum_u R_u^\star} \, , \forall u$ \tikzmark{label}\\
        Forward pass: $[\mb{\widehat{W}}_e, \mb{\widehat{\Omega}}_e, \widehat{\gamma}_e ] = \Phi_{\mathbf{\theta}}^e\left ( \mathbf{H_e^{(b)}}, \mb{R^{\star}} \right )$  \tikzmark{right} \\
        Loss: $\mathcal{L}_e \leftarrow \mathcal{L}(\mb{\widehat{W}}_e, \mb{\widehat{\Omega}}_e, \widehat{\gamma}_e,  \mb{R}^\star)$\\
    }
    Update parameters: 
    $\theta \leftarrow \theta - \lambda \cdot \sum_e \frac{1}{k} \nabla  \mathcal{L}_e$ \\
} \tikzmark{bottom}
}
\begin{tikzpicture}[overlay, remember picture]
    \draw [decorate,decoration={brace,amplitude=0.5em},thick,black] 
    ($(right)!(top1.north)!($(right)-(0,1)$)$) -- ($(right)!(bottom1.south)!($(right)-(0,1)$)$);

    \node[anchor=west, rotate=-90] at ($(label1) + (1.05,1)$) {\textbf{Pre-training}};
\end{tikzpicture}

\begin{tikzpicture}[overlay, remember picture]
    \draw [decorate,decoration={brace,amplitude=0.5em},thick,black] 
    ($(right)!(top.north)!($(right)-(0,1)$)$) -- ($(right)!(bottom.south)!($(right)-(0,1)$)$);

    \node[anchor=west, rotate=-90] at ($(label) + (3,1.1)$) {\textbf{Multi-objective}};
\end{tikzpicture}
\end{algorithm}

\subsection{Domain Adaptation}\label{sec:domain_adapt}

The pre-trained model consists of a shared feature extractor and $k$ branching output heads. For model deployment at a new site, we combine the feature extractor with a new site-specific output layer. 
To accommodate deployment with no adaptation data, we simply train offline a default output head with the entire training dataset. This default head is used for all deployment environments in the zero-shot case.

In few-shot scenarios, to prevent overfitting, we augment the adaptation data with similar training environments, selected via a similarity metric computed using the pre-trained feature extractor $f_{\mathbf{\theta}}\left ( \cdot \right )$.

Consider $l$ samples of \gls{CSI} data $\{\mb{H}_e^{(i)}\}_{i=1}^l$ from a site $e$. We denote its feature vector as $$\mathbf{\psi}_e = \frac{1}{l} \sum_{i=1}^{l} f_{\theta}(\mb{H}_e^{(i)}).$$ We can compute the cosine similarity pairwise between the feature vectors of the $e$-th training site and the deployment site $d$ as $$ D_{e, d} = \frac{\mathbf{\psi}_e \cdot \mathbf{\psi}_d}{\|\mathbf{\psi}_e\|\|\mathbf{\psi}_d\|}.$$ The augmented dataset is then generated by selecting the top $N$ similar training environments to the deployment site. 

%% file: Results.tex
\section{Numerical Results}

In this section, we evaluate the performance of our proposed model on three distinct datasets listed below. Note that these deployment sites were not included in the training dataset, to accurately assess the out-of-distribution generalization of the foundation model.
\begin{enumerate}
    \item \textit{Ericsson}: Industrial area which features wide streets and low buildings. 75\% of the users in this zone have a clear line-of-sight with the base station.
    \item \textit{Decarie}: Residential neighborhood characterized by three-story apartment buildings. Approximately half of the users have a line-of-sight with the base station.
    \item \textit{Sainte-Catherine}: Downtown area filled with sky-scrapers. 75\% of the simulated users do not have a line-of-sight with the base station. The high density of buildings in the area generates many signal reflections.
\end{enumerate}

\subsection{Deployment}
For each deployment environment introduced, we consider three scenarios with respect to data availability. When quoting the dataset sizes, a data sample corresponds to the full \gls{CSI} for one user position.
\begin{enumerate}
    \item \emph{Zero-shot:} We assess the zero-shot performance of our foundation model by deploying it with the default output head.
    \item \emph{Few-shot:} Here, we evaluate our data augmentation method. To compute the feature vector of the deployment site, we use $10$ CSI samples. The fine-tuning dataset is comprised of the top $N=5$ training environments most similar to the deployment site, where similarity is evaluated through the method presented in Section~\ref{sec:domain_adapt}.
    \item \emph{Full-dataset:} With access to an abundant deployment dataset, we can train the output layer using only local data, further adapting the model to the specific environment. We consider that a ``full'' deployment-site dataset consists of 5000 CSI samples.
\end{enumerate}

\begin{figure}[t]
    \centering
    \includegraphics[width=0.9\linewidth]{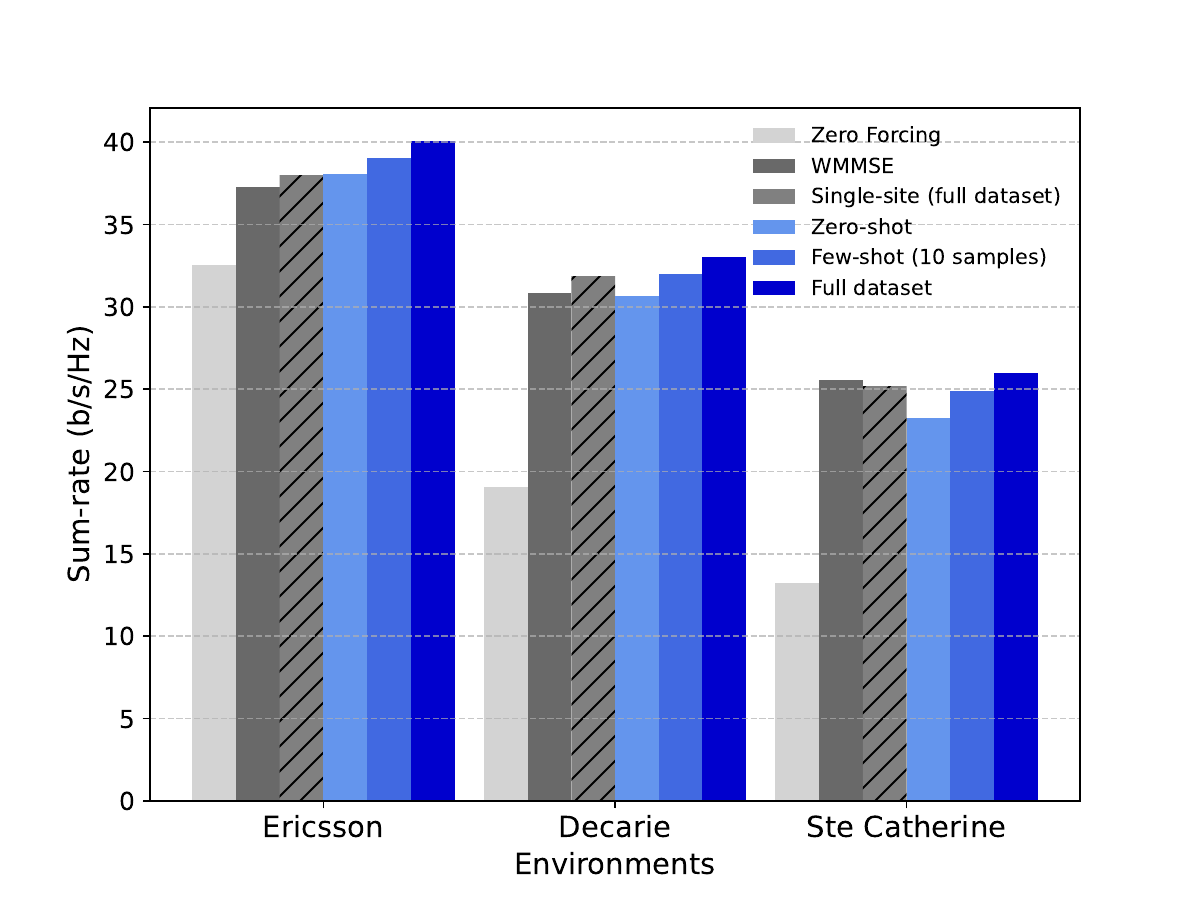}
    \caption{Maximum achievable sum-rate of the adapted foundation model in unseen deployment environments ($N_U=4$, $N_T=64$, $-\log\sigma^2 = 13$).}
    \label{fig:results}
\end{figure}

\begin{figure}[t]
    \centering
    \includegraphics[width=0.9\linewidth]{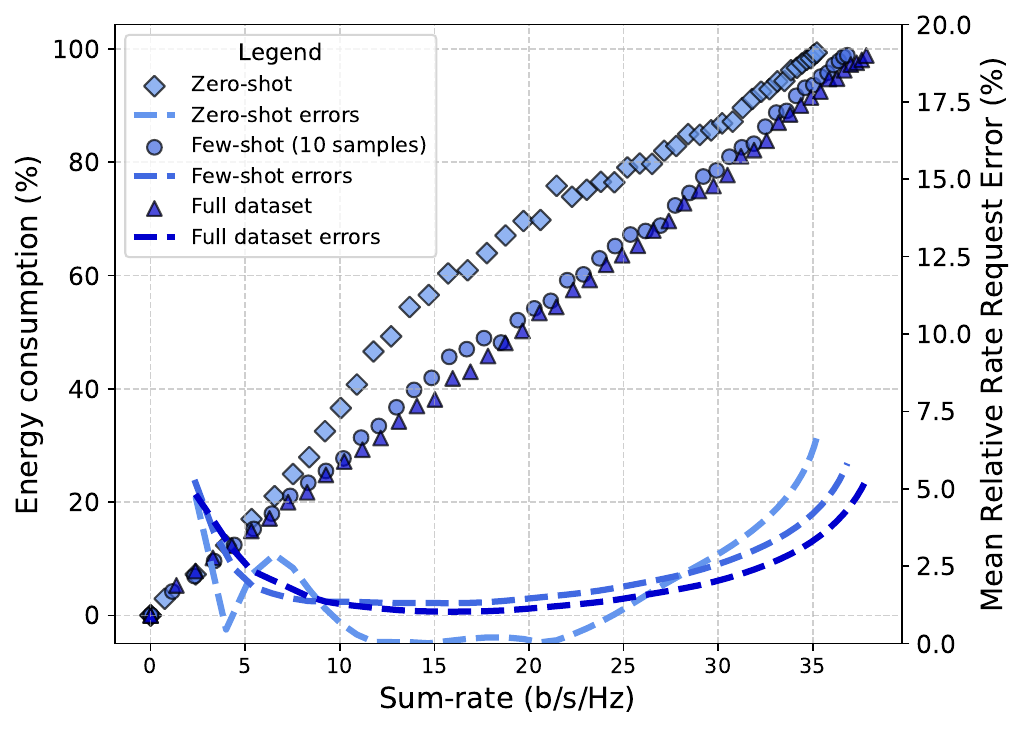}
    \caption{ Adaptive rate-power tradeoff on ``Ericsson'' site for three data availability settings ($N_U=4$, $N_T=64$, $-\log\sigma^2 = 13$).}
    \label{fig:tradeoff}
\end{figure}

In Figure~\ref{fig:results}, we present the results of our deployed foundation model for maximum sum-rate precoding - obtained by requesting very large user rates. In this setting, transmit power is maximal and all antennas are turned on.

On all deployment sites, zero-shot deployment of the pre-trained model outperforms \gls{ZF}. At the Ericsson site where users are mostly \gls{LOS}, near-optimal \gls{WMMSE} is surpassed. 

Few-shot adaptation further improves the achievable sum-rate, by ensuring that the site-specific output layer is trained with a dataset similar to the full deployment site data. While all three sites see an increase in performance from few-shot adaptation, it is most pronounced at the Sainte-Catherine \gls{BS} situated in a downtown setting. Designing a precoder to serve \gls{NLOS} users is generally more difficult, because of the rich scattering environment. These types of areas will most benefit from adaptation as the model can learn the unique characteristics of the propagation environment.

Adapting the foundation model using the full dataset further improves performance, surpassing \gls{WMMSE} on all sites. Interestingly, this approach outperforms a \gls{DL} model trained from scratch on the full single-site dataset. Fine-tuning not only reduces training complexity, but also achieves better performance by leveraging the rich priors learned.

\subsection{Adaptive Per-User Rate-Power Tradeoff}

To visualize the energy-throughput tradeoff, we simulate random user-rate requirements to be satisfied by the precoder, and record the energy cost for each transmission. Figure~\ref{fig:tradeoff} ranks the observations based on the average sum-rate. As the user-rate requirements decrease, the model learns to turn off certain antennas and reduce the transmit power in order to save energy. We provide the tradeoff curve for zero-shot, few-shot and full dataset adaptation. We can see that training data samples particularly improves the performance of the energy minimization task. 

On the same figure, we plot the mean relative user-rate request error, demonstrating that our models precisely match the requested rates, with an average relative error below 5\%.

\subsection{Complexity Analysis}
The number of \gls{FLOPs} needed to compute the precoder using the two baseline methods and the proposed model are summarized in Table~\ref{tbl:complexity}. 
The complexity for the three algorithms was determined by counting the number of real additions and multiplications. 
Since \gls{WMMSE} is iterative, the complexity is in terms of the number of iterations $I$, with $I=16$ corresponding to the average number of iterations over the three deployment sites. For the foundation model, we refer to the dimensions detailed in Table~\ref{tab:dims}.

\gls{ZF} remains the most efficient method since it only involves one matrix inversion, but achieves poor performance, especially in environments with a low \gls{SNR}.
\gls{WMMSE} on the other hand achieves high performance but requires inverting a matrix at each iteration of the optimization process. The proposed DNN model is able to outperform WMMSE in all scenarios (with adaptation) but with $8\times$ lower complexity.

\begin{table}[t]
    \centering
    \caption{Number of \gls{FLOPs} to compute the precoder ($N_U=4$, $N_T=64$)}
    \label{tbl:complexity}
    \resizebox{\columnwidth}{!}{%
    \begin{tabular}{@{}ccl@{}}
    \toprule
    \textbf{Algorithm} &
      \begin{tabular}[c]{@{}c@{}}\textbf{\# of FLOPs} \\ $\mathbf{\times 10^6}$\end{tabular} &
      \multicolumn{1}{c}{\textbf{Complexity expression}} \\ \midrule
    \multicolumn{1}{c|}{ZF} &
      \multicolumn{1}{c|}{0.014} &
      $7\left( \frac{2}{3}N_U^3 + 2N_U^2N_T\right )$ \\ \midrule
    \multicolumn{1}{c|}{WMMSE} &
      \multicolumn{1}{c|}{\begin{tabular}[c]{@{}c@{}}93.5\\ $(I=16)$\end{tabular}} &
      \begin{tabular}[c]{@{}l@{}}$ I \big( \frac{14}{3}N_UN_T^3 + 12N_U^2N_T^2 + 12N_U^2N_T + 9N_UN_T^2$ \\ $+ 8N_UN_T + 5N_U^2 + \frac{68}{3}N_U \big)$\end{tabular} \\ \midrule
    \multicolumn{1}{c|}{Proposed} &
      \multicolumn{1}{c|}{10.8} &
    $\left ( 2^{19} + 2^{21}\right )N_U + 2048N_U^2 + 1024N_UN_T$\\
\bottomrule
    \end{tabular}%
}
\end{table}

\begin{table}[t]
  \centering
  \caption{Dimensions and Hyperparameters of the foundation model.}
  \label{tab:dims}
  \resizebox{\columnwidth}{!}{%
  \begin{tabular}{@{}cclc@{}}
    \toprule
    \textbf{Hyperparameter} & \textbf{Value} & \textbf{Hyperparameter} & \textbf{Value} \\
    \midrule
    \multicolumn{1}{c|}{Token dimension}         & \multicolumn{1}{c|}{$2 \cdot N_T$}  & \multicolumn{1}{c|}{Embedding dimension}     & 128            \\
    \multicolumn{1}{c|}{Sequence length}         & \multicolumn{1}{c|}{$N_U + 1$}      & \multicolumn{1}{c|}{FFN hidden dimension}     & 1024           \\
    \multicolumn{1}{c|}{Number of heads}         & \multicolumn{1}{c|}{2}              & \multicolumn{1}{c|}{Dropout}                 & 0.05           \\
    \multicolumn{1}{c|}{Number of layers}        & \multicolumn{1}{c|}{4}              & \multicolumn{1}{c|}{Batch size}      & 1000         \\
    \multicolumn{1}{c|}{Optimizer}               & \multicolumn{1}{c|}{Adam}          & \multicolumn{1}{c|}{Learning rate}            & $10^{-4}$      \\
    \bottomrule
  \end{tabular}%
  }
\end{table}

%% file: Conclusion.tex
\section{Conclusion}

\gls{DL} will be key in enabling energy-efficient \gls{mMIMO} technology thanks to its ability to learn site-specific low-complexity approximations. Yet, challenges related to data availability and adaptation still need to be addressed before deploying \gls{DL}-based solutions in practice. In this paper, we proposed training a foundation model for \gls{mMIMO} precoding with an adaptive per-user rate-power tradeoff. The pre-trained general feature extractor can be transferred to new deployment sites directly, reducing the need for training data and improving the efficiency of adaptation. We show excellent performance of our model with no/few training data samples at deployment sites. Additionally, the computational complexity of our approach is greatly inferior to optimization based methods like \gls{WMMSE}.

%% file: main.bbl
\begin{thebibliography}{10}
\providecommand{\url}[1]{#1}
\csname url@samestyle\endcsname
\providecommand{\newblock}{\relax}
\providecommand{\bibinfo}[2]{#2}
\providecommand{\BIBentrySTDinterwordspacing}{\spaceskip=0pt\relax}
\providecommand{\BIBentryALTinterwordstretchfactor}{4}
\providecommand{\BIBentryALTinterwordspacing}{\spaceskip=\fontdimen2\font plus
\BIBentryALTinterwordstretchfactor\fontdimen3\font minus \fontdimen4\font\relax}
\providecommand{\BIBforeignlanguage}[2]{{%
\expandafter\ifx\csname l@#1\endcsname\relax
\typeout{** WARNING: IEEEtran.bst: No hyphenation pattern has been}%
\typeout{** loaded for the language `#1'. Using the pattern for}%
\typeout{** the default language instead.}%
\else
\language=\csname l@#1\endcsname
\fi
#2}}
\providecommand{\BIBdecl}{\relax}
\BIBdecl

\bibitem{iui}
T.~L. Marzetta, ``{Noncooperative Cellular Wireless with Unlimited Numbers of Base Station Antennas},'' \emph{IEEE Trans. on Wireless Communications}, vol.~9, no.~11, pp. 3590--3600, 2010.

\bibitem{us_dl_mm_bf}
H.~Hojatian, J.~Nadal, J.-F. Frigon, and F.~Leduc-Primeau, ``Unsupervised deep learning for massive {MIMO} hybrid beamforming,'' \emph{IEEE Trans. on Wireless Communications}, vol.~20, no.~11, pp. 7086--7099, 2021.

\bibitem{slmp}
A.~G. Pathapati, N.~Chakradhar, P.~Havish, S.~A. Somayajula, and S.~Amuru, ``Supervised deep learning for {MIMO} precoding,'' in \emph{2020 IEEE 3rd 5G World Forum (5GWF)}, 2020, pp. 418--423.

\bibitem{low_complexity_dl}
M.~Zhang, J.~Gao, and C.~Zhong, ``A deep learning-based framework for low complexity multiuser {MIMO} precoding design,'' \emph{IEEE Trans. on Wireless Communications}, vol.~21, no.~12, pp. 11\,193--11\,206, 2022.

\bibitem{unsupervised_graph}
Y.~Zhang, J.~Yang, Q.~Liu, Y.~Liu, and T.~Zhang, ``Unsupervised learning-based coordinated hybrid precoding for {MmWave} massive {MIMO}-enabled {HetNets},'' \emph{IEEE Trans. on Wireless Communications}, vol.~23, no.~7, pp. 7200--7213, 2024.

\bibitem{maml_hbf}
M.~Sun, S.~Wu, H.~Wang, and S.~Yang, ``Adaptive massive {MIMO} hybrid precoding based on meta learning,'' in \emph{2023 Int. Conf. on Wireless Communications and Signal Processing (WCSP)}, 2023, pp. 189--194.

\bibitem{mltl}
Y.~Yuan, G.~Zheng, K.-K. Wong, B.~Ottersten, and Z.-Q. Luo, ``Transfer learning and meta learning-based fast downlink beamforming adaptation,'' \emph{IEEE Trans. on Wireless Commun.}, vol.~20, no.~3, pp. 1742--1755, 2021.

\bibitem{sage}
A.~H. Karkan, H.~Hojatian, J.-F. Frigon, and F.~Leduc-Primeau, ``{SAGE-HB}: Swift adaptation and generalization in massive {MIMO} hybrid beamforming,'' in \emph{2024 IEEE International Conference on Machine Learning for Communication and Networking (ICMLCN)}, 2024, pp. 323--328.

\bibitem{catak2025bert4mimofoundationmodelusing}
\BIBentryALTinterwordspacing
F.~O. Catak, M.~Kuzlu, and U.~Cali, ``{BERT4MIMO}: A foundation model using {BERT} architecture for massive {MIMO} channel state information prediction,'' 2025. [Online]. Available: \url{https://arxiv.org/abs/2501.01802}
\BIBentrySTDinterwordspacing

\bibitem{foundation_2}
Z.~Liu, L.~Wang, L.~Xu, and Z.~Ding, ``Deep learning for efficient {CSI} feedback in massive {MIMO}: Adapting to new environments and small datasets,'' \emph{IEEE Trans. on Wireless Communications}, vol.~23, no.~9, pp. 12\,297--12\,312, 2024.

\bibitem{noisy_csi}
H.~Hojatian, Z.~Mlika, J.~Nadal, J.-F. Frigon, and F.~Leduc-Primeau, ``Learning energy-efficient transmitter configurations for massive {MIMO} beamforming,'' \emph{IEEE Trans. on Machine Learning in Communications and Networking}, vol.~2, pp. 939--955, 2024.

\bibitem{zf}
M.~A. Albreem, A.~H.~A. Habbash, A.~M. Abu-Hudrouss, and S.~S. Ikki, ``Overview of precoding techniques for massive {MIMO},'' \emph{IEEE Access}, vol.~9, pp. 60\,764--60\,801, 2021.

\bibitem{shi_wmmse}
Q.~Shi, M.~Razaviyayn, Z.-Q. Luo, and C.~He, ``An iteratively weighted mmse approach to distributed sum-utility maximization for a {MIMO} interfering broadcast channel,'' \emph{IEEE Trans. on Signal Processing}, vol.~59, no.~9, pp. 4331--4340, 2011.

\bibitem{OpenStreetMap}
{OpenStreetMap contributors}, ``{Planet dump retrieved from https://planet.osm.org },'' \url{ https://www.openstreetmap.org }, 2017.

\end{thebibliography}
